\documentclass{elsarticle}
\makeatletter
\def\ps@pprintTitle{%
 \let\@oddhead\@empty
 \let\@evenhead\@empty
 \def\@oddfoot{\centerline{\thepage}}%
 \let\@evenfoot\@oddfoot}
\makeatother

\usepackage[utf8]{inputenc}
\usepackage{amsmath, german}
\usepackage{graphicx, xcolor, tikz, url}
\usepackage[top=2cm, bottom=2cm, right=2cm, left=2cm]{geometry}

\usepackage[FIGTOPCAP,nooneline]{subfigure} 
\usetikzlibrary{fit}

\renewcommand{\phi}{\varphi}
\renewcommand{\theta}{\vartheta}

\newcommand{\Chi}{\mathcal{X}}

\renewcommand{\bar}[1]{\overline{#1}}
\newcommand{\abs}[1]{\left|#1\right|}

\newcommand{\lcb}{\left\{}

\newcommand{\rcb}{\right\}}

\newcommand{\lpm}{\begin{pmatrix}}
\newcommand{\rpm}{\end{pmatrix}}

\begin{document}
\title{Berechnung einer lokalen COVID--19 Reproduktionszahl für das nördliche Rheinland--Pfalz \newline ----- \newline 
Calculation of a local COVID--19 reproduction number for the northern Rhineland--Palatinate}
\author[1]{Thomas Götz\corref{cor1}}
\ead{goetz@uni-koblenz.de}
\author[1]{Silja Mohrmann}
\ead{siljamohrmann@uni-koblenz.de}
\author[1]{Robert Rockenfeller}
\ead{rrockenfeller@uni-koblenz.de}
\author[1]{Moritz Schäfer}
\ead{moritzschaefer@uni-koblenz.de}
\author[1]{Karunia Putra Wijaya}
\ead{karuniaputra@uni-koblenz.de}

\address[1]{Mathematical Institute, University of Koblenz--Landau, 56070 Koblenz, Germany}

\cortext[cor1]{Corresponding author.}

\begin{abstract}
Seit dem Beginn der Corona--Pandemie Anfang März 2020 werden für Deutschland neben den täglichen Infektionszahlen (Neuinfektionen sowie Gesamtinfektionen) verschiedene Parameter zur Beschreibung der Ausbreitung angegeben, welche auch für politische Entscheidungen herangezogen werden. Hierzu zählen neben der Übersterblichkeit und der $7$--Tage--Inzidenz auch die Verdopplungszeit $T_2$ sowie die Reproduktionszahl $R_t$. Zu Letzterer finden sich auf der Internetseite des Robert--Koch--Institutes~\cite{EstR:RKI} verschiedene Schätzungen, welche aus den Fallzahlen für ganz Deutschland berechnet werden; lokale Unterschiede werden hier nicht berücksichtigt. In dem vorliegenden Artikel werden die Berechnungen des RKI auf einer lokalen Ebene untersucht am Beispiel des nördlichen Rheinland-Pfalz und dessen Landkreisen. Hierbei wird nicht das Melde-, sondern das Erkrankungsdatum als Referenz für die Berechnung von $R_t$ herangezogen. Für Fälle, bei denen das Erkrankungsdatum nicht bekannt ist, wird zunächst eine angepasste verallgemeinerte Extremwertverteilung (GEV) auf die Daten, bei denen die Angabe des Meldeverzugs (Differenz zwischen Krankheitsbeginn und Meldedatum) vorhanden ist, angepasst und auf weitere Merkmale wie lokale sowie demografische Differenzen untersucht. Diese GEV--Verteilung wird dann zur Berechnung der Meldeverzüge der unvollständigen Datenpunkte verwendet. Bei der Berechnung des täglichen Wertes von $R_t$ zwischen Ende Februar und Ende Oktober konnte ein im Vergleich mit den deutschlandweiten Zahlen ähnlicher Verlauf der Reproduktionszahl festgestellt werden. Erwartbar größeren statistischen Schwankungen wurden im Sommer, vor allem durch geringere Fallzahlen, festgestellt, wobei die Werte für das nördliche Rheinland--Pfalz seit etwa Mitte September konstant über $1$ liegen. Die Berechnungen können ebenso auf weitere Regionen und Landkreise übertragen werden.

\smallskip

-----
\smallskip

Since the beginning of the corona pandemic in March 2020, various parameters for describing the spread of the disease have been specified for Germany in addition to the daily infection figures (new infections and total infections), which are also used for political decisions. In addition to excess mortality and the weekly incidence, these include the doubling time $T_2$ and the reproduction number $R_t$. For the latter, various estimates can be found on the website of the Robert--Koch--Institute, see~\cite{EstR:RKI}, which are calculated from the case numbers for all of Germany; local differences are not taken into account here. In the present article, the calculations of the RKI on a local level are examined using the example of northern Rhineland--Palatinate and its districts. Here, not the reporting date but the onset of illness is used as a reference for the calculation of $R_t$. For cases where the onset of illness is not known, an adjusted generalized extreme value distribution (GEV) is first fitted to the data for which the reporting delay (difference between the onset of illness and the reporting date) is available and examined for further characteristics such as local as well as demographic differences. This GEV distribution is then used to calculate the reporting delays of incomplete data points. The calculation of the daily value of $R_t$ between the end of February and the end of October showed a similar course of the reproductive rate compared to the nationwide figures. Expectably larger statistical fluctuations were observed in the summer, mainly due to lower case numbers. The values for northern Rhineland--Palatinate have been consistently above $1$ since about mid--September. The calculations can also be transferred to other regions and administrative districts.
\end{abstract}
\maketitle

%\todo{Grundsätzliche Frage (Robert): Alle Dezimalzahlen mit Punkt oder Komma?}
%\TG{Punkt ist für mich ok.}
%
%
\section{Datenlage}
\begin{figure}
    \centering
    \subfigure[]{
    \includegraphics[width=\textwidth]{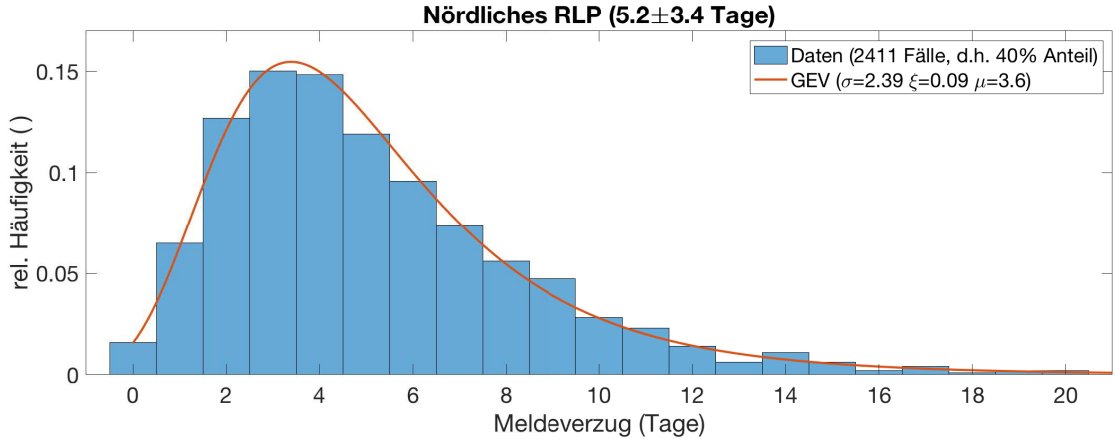} \label{F:GEV_RLP}}
    
    \subfigure[]{
    \hspace{-2.42cm}
    \includegraphics[width=1.241\textwidth]{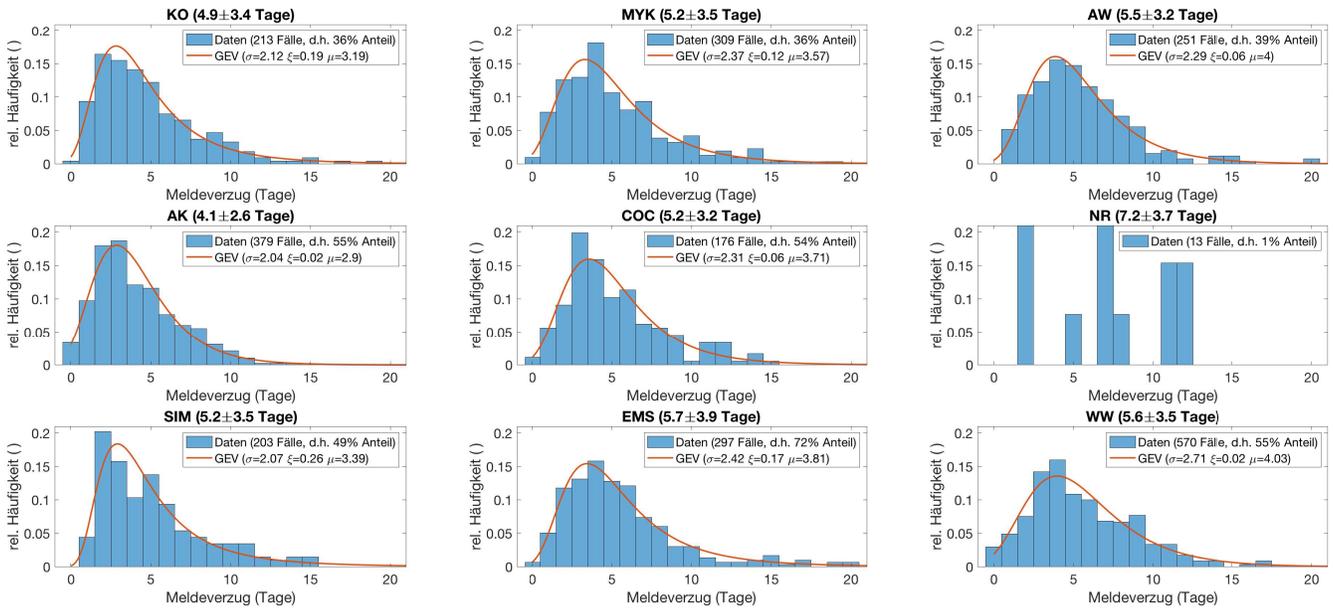} \label{F:GEV_LK}}
    \caption{Meldeverzug \subref{F:GEV_RLP} im gesamten nördlichen Rheinland--Pfalz und \subref{F:GEV_LK} aufgeteilt in die Stadt Koblenz (KO) und die Landkreise Mayen--Koblenz (MYK), Ahrweiler (AW), Altenkirchen (AK), Cochem--Zell (COC), Neuwied (NR), Rhein--Hunsrück--Kreis (SIM), Rhein--Lahn--Kreis (EMS) und Westerwald (WW). Dargestellt ist jeweils das Histogramm für den Meldeverzug sowie eine angepasste Wahrscheinlichkeitsverteilung (verallgemeinerte Extremwertverteilung / GEV--distribution, siehe \ref{A:Details}). Mittelwerte plus/minus Stichprobenstandardabweichungen der Meldeverzüge sind, ebenso wie der Anteil der Meldeverzüge an der Gesamtzahl der Datensätze, in Klammern angegeben.
        }
    \label{F:Meldeverzug}
\end{figure}

Das Robert--Koch--Institut (RKI) veröffentlicht täglich aktualisierte Daten\-sätze zu den gemeldeten COVID--19 Infektionen~\cite{COV:RKI}. Für jede gemeldete Infektion enthält der Datensatz neben demographischen Angaben (Alter, Geschlecht) auch geographische Informationen (Bundesland und Landkreis) sowie Angaben zum Meldedatum (wann wurde der Fall beim jeweiligen Gesundheitsamt registriert?), für Details siehe~\cite{COV_readme:RKI}. Weniger als die Hälfte der Datensätze (Stand 31. Oktober ca.~$40\%$ im gesamten nördlichen Rheinland--Pfalz) enthält zusätzliche Informationen zum vermutlichen Datum des Erkrankungsbeginns. In Abbildung~\ref{F:Meldeverzug} wird der Meldeverzug (entspricht der Differenz zwischen Erkrankungsbeginn und Meldedatum) für die im nördlichen Rheinland--Pfalz gemeldeten Fälle dargestellt, ebenso für die Stadt Koblenz sowie für die acht Landkreise (Mayen--Koblenz, Ahrweiler, Altenkirchen, Cochem--Zell, Neuwied, Rhein--Hunsrück--Kreis, Rhein--Lahn--Kreis und Westerwaldkreis).

\section{Berechnung der lokalen Reproduktionszahl}
Für Datensätze ohne Angabe des Erkrankungsbeginns kann dieser gemäß der angepassten GEV--Verteilung geschätzt werden. Basierend hierauf wird analog zu der in~\cite{EstR:RKI} beschriebenen Vorgehensweise die aktuelle Reproduktionszahl $R_t$ berechnet; Details hierzu beschreiben wir in~\ref{A:Details}. Um statistische Schwankungen teilweise zu eliminieren, verwenden wir eine Mittelung der Anzahl der Erkrankungsbeginne über ein Zeitintervall von $\tau=4$ bzw.~$\tau=7$ Tagen und geben entsprechend die aktuelle, über $4$ bzw.~$7$ Tage gemittelte Reproduktionszahl $R_{4,t}$ bzw.~$R_{7,t}$ an. Hierbei ist zu beachten, dass die Werte für beide Reproduktionszahlen durch die maximal möglichen Meldeverzüge von 21 Tagen noch in ebendiesem Zeitraum veränderlich bleiben, sodass die jüngsten, aktuellen Resultate gesondert dargestellt werden.

\begin{figure}
    \centering
    \includegraphics[width=\textwidth]{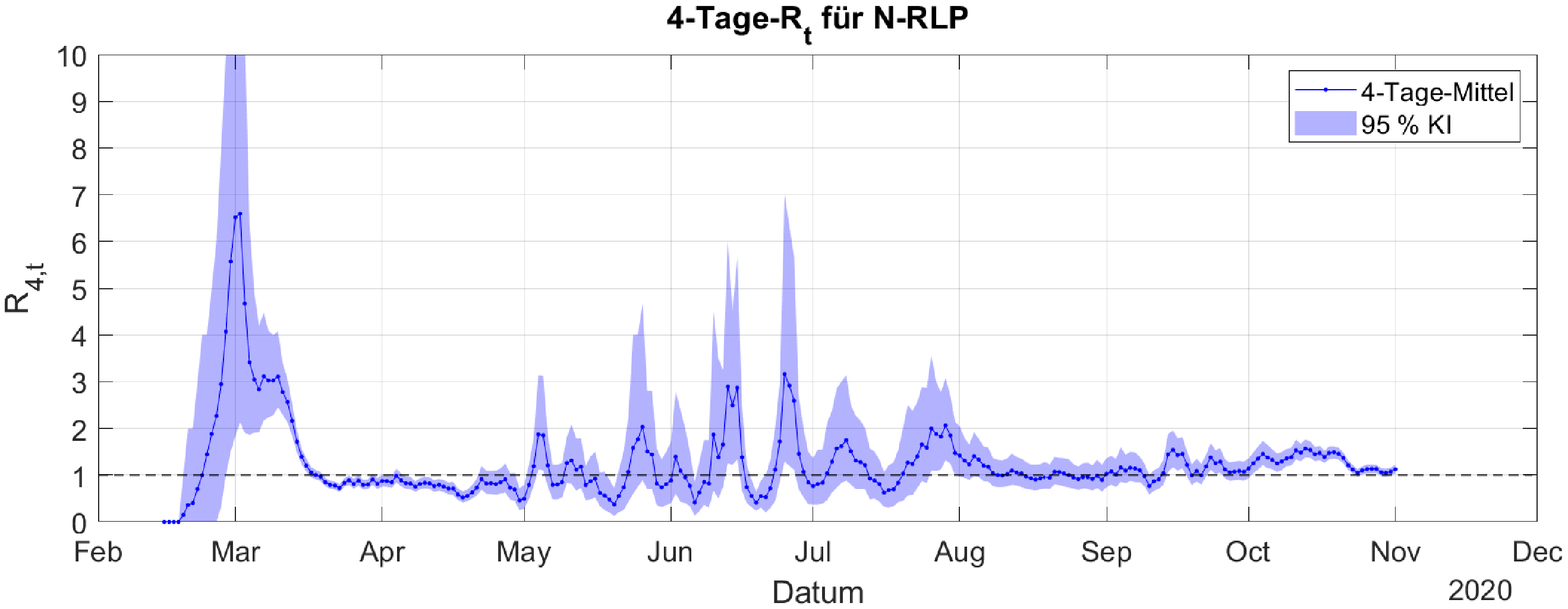}
    \caption{Reproduktionszahl $R_{4,t}$, gemittelt über $4$ Tage, für das gesamte nördliche Rheinland--Pfalz. Dargestellt ist der Mittelwert sowie das $95\%$--Konfidenzintervall basierend auf $1000$ Realisierungen der Schätzung des Meldeverzugs.}
    \label{F:R4-nRLP}
\end{figure}
\begin{figure}
    \centering
    \includegraphics[width=\textwidth]{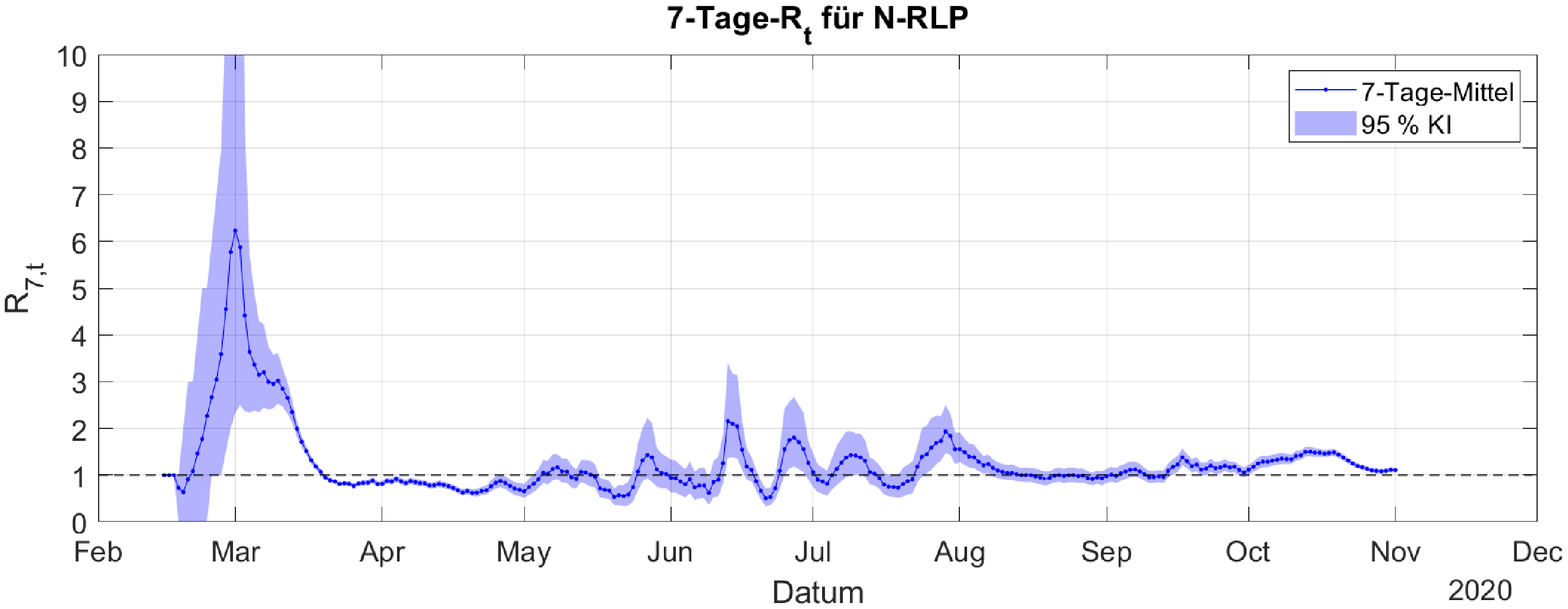}
    \caption{Reproduktionszahl $R_{7,t}$, gemittelt über $7$ Tage, für das gesamte nördliche Rheinland--Pfalz. Dargestellt ist der Mittelwert sowie das $95\%$--Konfidenzintervall basierend auf $1000$ Realisierungen der Schätzung des Meldeverzugs.}
    \label{F:R7-nRLP}
\end{figure}
\begin{figure}
    \centering
    \includegraphics[width=\textwidth]{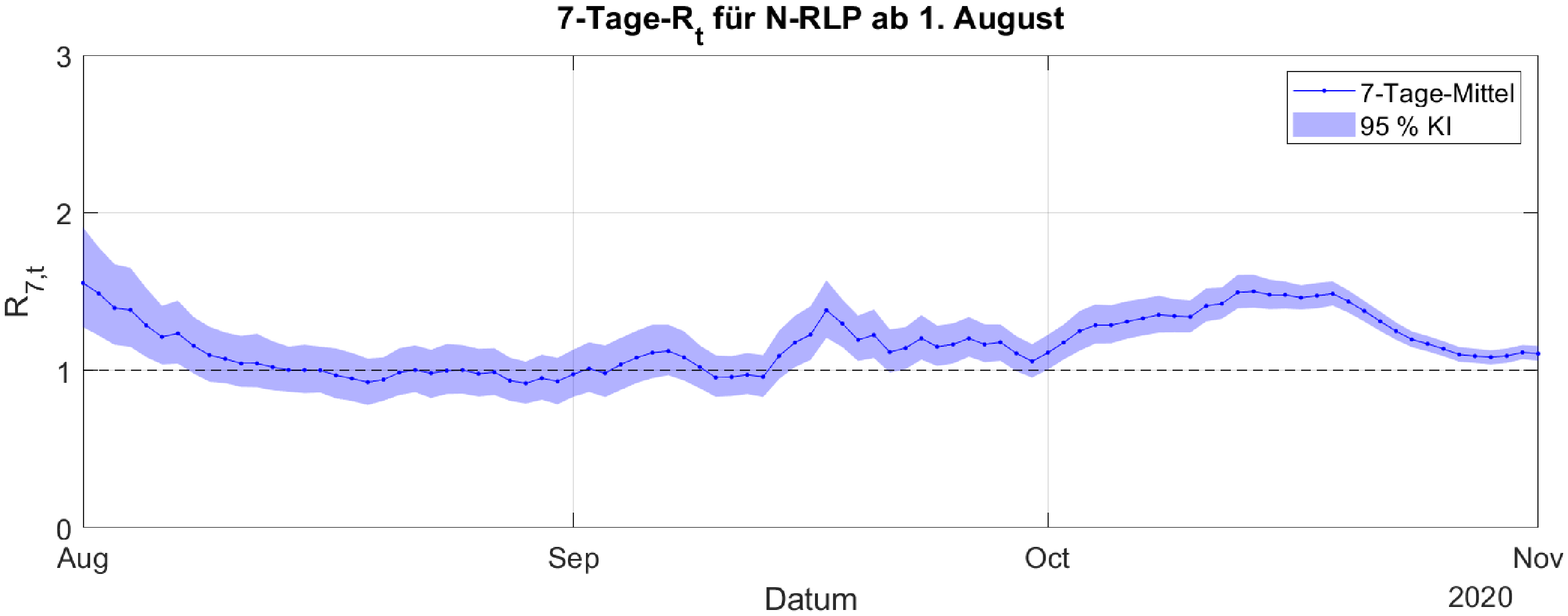}
    \caption{Reproduktionszahl $R_{7,t}$, gemittelt über $7$ Tage, für das gesamte nördliche Rheinland--Pfalz seit dem 1. August. Dargestellt ist der Mittelwert sowie das $95\%$--Konfidenzintervall basierend auf $1000$ Realisierungen der Schätzung des Meldeverzugs.}
    \label{F:R7a-nRLP}
\end{figure}

\section{Bewertung der Resultate}

Betrachtet man die Daten zum Meldeverzug in den einzelnen Landkreisen des nördlichen Rheinland--Pfalz, vgl.~Abbildung~\ref{F:Meldeverzug}, so fällt auf, dass für den Landkreis Neuwied in lediglich $1\%$ aller Fälle ($13$ von gesamt 991) der Erkrankungsbeginn verzeichnet ist. Eine Erklärung hierfür, bzw. eine Rückmeldung des zuständigen Gesundheitsamtes fehlen bislang. In den anderen Landkreisen liegt für jeweils zwischen $39\%$ (Koblenz und Mayen--Koblenz) und $72\%$ (Rhein--Lahn--Kreis) der Fälle sowohl das Meldedatum als auch der vermutete Erkrankungsbeginn vor. Der mittlere Meldeverzug als Differenz zwischen Erkrankungsbeginn und Meldedatum schwankt zwischen $4.1$ Tagen (Altenkirchen) und $5.7$ Tagen (Rhein--Lahn--Kreis). Im Durchschnitt beträgt der Meldeverzug für das gesamte nördliche Rheinland--Pfalz etwa $5.2$ Tage. Die jeweilige Verteilung des Meldeverzug verhält sich für alle Landkreise (mit Ausnahme von Neuwied aus den genannten Gründen) ähnlich.

Bei der Schätzung der $7$--Tage Reproduktionszahl für das nördliche Rheinland--Pfalz (Grafik~\ref{F:R7-nRLP}) fällt auf, dass zu Beginn der Pandemie (März) die Reproduktionszahl teils sehr große Werte annahm. Dies entspricht dem damaligen raschen Wachstum mit Verdopplungszeiten von minimal $2$--$3$ Tagen. Die Einführung der Kontaktbeschränkungen und Einschränkungen im öffentlichen Leben ab Mitte März reduzierte den $R$--Wert in der Folge unterhalb der kritischen Marke von $1$. Die Oszillationen im Sommer (Juni-Juli) lassen sich teilweise als statistische Schwankungen aufgrund der niedrigen Fallzahlen erklären (es wurden insgesamt nur ca. 220 Krankheitsbeginne im gesamten nördlichen Rheinland-Pfalz in diesem Zeitfenster simuliert). Die niedrigen Fallzahlen während des Sommers verursachen auch die teilweise extrem großen Schwankungen des $4$--Tage $R$--Wertes für die einzelnen Landkreise. In Perioden, in denen es überhaupt keine Neuinfektionen innerhalb eines Landkreises gab, z.b. Mitte Juni--Mitte Juli im Rhein--Hunsrück--Kreis, siehe Abbildungen~\ref{F:R4-SIM}--\ref{F:R7-SIM}, lässt sich \emph{kein} $R$--Wert berechnen. Ab September ist ein sukzessives Ansteigen des $R$--Wertes konstant oberhalb der Marke von $1$ (bis auf wenige Ausnahmen) zu beobachten. Dieser Anstieg hat sich im Oktober konsolidiert und wurde für den 10.~Oktober auf $1.33$ ($1.23$---$1.44$) geschätzt. Ein $R$--Wert von $1.33$ entspricht in etwa einer Verdopplungszeit von 3 Wochen.

\clearpage
\appendix

\section{Mathematische Details} \label{A:Details}

Der Datensatz des RKI enthält zum Stichtag 31. Oktober 2020 insgesamt $N=5954$ Datensätze aus dem nördlichen Rheinland--Pfalz, d.h. aus der Stadt Koblenz (600 Datensätze) und den Landkreisen Mayen--Koblenz (851), Ahrweiler (636), Altenkirchen (692), Cochem--Zell (325), Neuwied (991), Rhein--Lahn--Kreis (410), Rhein--Hunsrück--Kreis (414) und Westerwaldkreis (1035). Für jeden dieser Datensätze bezeichne $T_{M,i}$, $i=1,\dots, N$, das Meldedatum. Für die Teilmenge $j\in \lcb i_1,\dots, i_m \rcb \subset \lcb 1,\dots N\rcb$ von Fallindizes ist auch das Datum $T_{K,j}$ des Erkrankungsbeginns bekannt. Der Meldeverzug $d_j=T_{M,j}-T_{K,j}$, d.h.~die Differenz zwischen dem Erkrankungsbeginn und der Registrierung des Infektionsfalls beim zuständigen Gesundheitsamt, kann als eine diskrete Zufallsvariable angesehen werden. Ein \emph{Pearson's $\Chi^2$--Verteilungstest} (21 Freiheitsgrade, 5\% Signifikanzniveau) mit sämtlichen in \emph{MatLab} (Mathworks, Natick, USA, Version 2020a, Statistical Toolbox) vorimplementierten Wahrscheinlichkeitsverteilungen ergab lediglich bei der \emph{verallgemeinerten Extremwertverteilung} (generalized extreme value distribution, GEV) keine Ablehnung der Nullhypothese $H_0(F):$ {\glqq}Die Meldeverzüge für das nördliche Rheinland-Pfalz genügen der Verteilung $F${\grqq}.
Die Wahrscheinlichkeitsdichte der GEV mit einen \emph{scale}--Parameter $\sigma$, einen \emph{shape}--Parameter $\xi$ und einen \emph{location}--Parameter $\mu$ ist wie folgt definiert~\cite{Jen55}:
\begin{equation}
    f_{GEV}(x) =\begin{cases} \exp\left(-\frac{x-\mu}{\sigma} \right)\cdot\exp\left(-\exp\left(-\frac{x-\mu}{\sigma} \right)\right) &\text{ , falls } \xi=0\\
    (1+\xi \cdot \frac{x-\mu}{\sigma})^{-(1+\frac 1 \xi)}\cdot \exp \left(-(1+\xi \cdot \frac{x-\mu}{\sigma})^{-\frac 1 \xi} \right)&\text{ , falls } \xi \neq 0 \text{ und } \xi\cdot \frac{x-\mu}{\sigma}>-1\\
    0 &\text{ , sonst}
    \end{cases} \;.
\end{equation}
 Die Verteilung der Meldeverzüge der einzelnen Landkreise (außer NR und WW) genügen ebenfalls einer GEV, siehe Fig.~\ref{F:Meldeverzug} und \ref{A:Kohorten}. Einzig für die Landkreise AW und SIM könnten die Meldeverzüge auch anderen Verteilungen genügen; beispielsweise Log--Logistisch, Log--Normal, Invers Normal oder Birnbaum--Sauders. 
Aufgrund dieses eindeutigen Ergebnisses verwenden wir jedoch die GEV als plausibelste Verteilung.

Für die Datensätze $i\in \lcb 1,\dots, N\rcb \setminus \lcb i_1, \dots, i_M\rcb$, für die \emph{kein} Erkrankungsbeginn vorliegt, ziehen wir eine Zufallsvariable $\tilde{d}_i\sim GEV(\sigma, \xi, \mu)$ für den Meldeverzug für jeden Landkreis separat bzw. für das gesamte nördliche Rheinland-Pfalz gemäß den gemäß der verallgemeinerten Extremwertverteilung mit den Werten aus Abbildung \ref{F:Meldeverzug}. Der maximale Meldeverzug liegt bei 21 Tagen; Werte in der Verteilung, die diesen Wert überschreiten, werden als 21 Tage festgelegt. Der geschätzte Erkrankungsbeginn ergibt sich dann zu $\tilde{T}_{K,i} = T_{M,i}-\tilde{d}_i$. Durch diese statistische Rekonstruktion der fehlenden Erkrankungbeginne basierend auf der geschätzten Verteilung des Meldeverzugs erhalten wir für \emph{alle} Fälle einen möglichen Erkrankungsbeginn. Mit $E_t=\abs{\lcb i: \tilde{T}_{K,i}=t\rcb}$ bezeichnen wir die Anzahl aller Fälle im Beobachtungsgebiet mit Erkrankungsbeginn am Tag $t$.

Die aktuelle Reproduktionszahl $R_t$ zum Zeitpunkt $t$ ist definiert als die mittlere Anzahl von Sekundärinfektionen, welche ein am Tag $t$ Infizierter verursacht. Geht man von einem seriellen Intervall $\sigma$ von $4$ Tagen aus, vgl.~\cite{EstR:RKI}, so ergibt sich die aktuelle Reproduktionszahl $R_t$ als das Verhältnis der heutigen neuen Erkrankungsfälle $E_t$ zu denen von vor $\sigma$ Tagen, also $R_t = \frac{E_t}{E_{t-\sigma}}$. Um die statistischen Schwankungen in den täglichen Neuerkrankungen teilweise zu eliminieren, verwenden wir anstelle der tagesaktuellen Werte ein gleitendes Mittel der letzten $\tau$ Tage, vergleiche auch~\cite{EstR:RKI}:
\begin{equation}
    R_{\tau, t} = \frac{\frac{1}{\tau}\sum_{k=0}^{\tau-1} E_{t-k}}{\frac{1}{\tau}\sum_{k=0}^{\tau-1} E_{t-\sigma-k}}
        = \frac{\bar{E}_{\tau,t}}{\bar{E}_{\tau,t-\sigma}}\;
\end{equation}
Hierbei bezeichnet $\bar{E}_{\tau,t} := \frac{1}{\tau}\sum_{k=0}^{\tau-1} E_{t-k}$ das gleitende Mittel der Neuerkrankungen der letzten $\tau$ Tage. Ergebnisse hierzu für das nördliche Rheinland--Pfalz mit $\tau=4$ sowie $\tau=7$ sind in Abbildungen~\ref{F:R4-nRLP} und \ref{F:R7-nRLP} dargestellt. Der detailliertere Verlauf des $R_{7,t}$--Wertes für das nördliche Rheinland-Pfalz seit Anfang August findet sich in Abbildung~\ref{F:R7a-nRLP} .

%\todo{@Silja: serielles Intervall $t\sim \log\calN$ als Log--Normalverteilung}

Die Reproduktionszahl $R_t$ steht in direktem Zusammenhang mit der \emph{Verdopplungszeit} $T_2$ der Infektionsfälle. In einer Bevölkerung, die noch größtenteils suszeptibel für die Erkrankung ist, lässt sich die Anzahl der aktuell Infizierten näherungsweise durch die Differentialgleichung
\begin{equation*}
    I' = \beta \frac{SI}{N} - \gamma I
\end{equation*}
beschreiben, vgl.~\cite{GH20,CDea20}. Für $S\sim N$ und $R_t\sim \frac{\beta}{\gamma}$ erhalten wir $I(t)=I_0 e^{\gamma(R_t-1)t}$. Hieraus ergibt sich die Verdopplungszeit zu
\begin{equation}
    T_2 = \frac{\log 2}{\gamma (R_t-1)}.
\end{equation}
Nimmt man eine Genesungsrate $\gamma=1/10$ an ---dies entspricht einer mittleren Infektionsdauer von $10$ Tagen, vgl.~\cite{CDea20}--- so ergibt sich ein direkter Zusammenhang zwischen der Reproduktionszahl $R_t$ und der Verdopplungszeit $T_2$.

\section{Kohortenanalyse} \label{A:Kohorten}

\subsection{Verteilungstest}

Um zu überprüfen, ob ein Datensatz einer gegebenen Verteilung genügt, wird ein Pearson's $\Chi^2$--Verteilungstest durchgeführt. Die Nullhypothese lautet $H_0(F):$ {\glqq}Die Daten genügen der Verteilungsfunktion $F${\grqq} und wird überprüft, indem der Prüfgröße 
$$\Chi^2_{F} := \sum\limits_{k=1}^N \frac{(O_k-E(F)_k)^2}{E(F)_k} $$
die Testgröße $\Chi^2_{1-\alpha,DOF}$ gegenübergestellt wird. Hierbei bezeichnet $N$ die Anzahl der Datenpunkte, $O_k$ deren beobachtete (observed) Ausprägung, $E(F)_k$ die aufgrund der Verteilung zu erwartende (expected) Ausprägung, $\alpha$ das Signifikanzniveau und $DOF$ ($=1-|\text{Kategorien}|$) die Anzahl der Freiheitsgrade.

Für die Meldeverzüge betrachten wir eine Höchstanzahl von 21 Tagen als zulässig, daher hat ein Test der Meldeverzüge auf GEV Verteilung 21 Freiheitsgrade (null Meldetage sind auch möglich). Bei einem Signifikanzniveau von $\alpha=5\%$ ist die Testgröße $\Chi^2_{0.95,21}=32.67$. Die Prüfgrößen der einzelnen Landkreise sind:
\begin{center}
\begin{tabular}{c||c|c|c|c|c|c|c|c|c||c}
Region 	& KO & MYK & AW & AK & COC & NR & SIM & EMS & WW & Nord--RLP \\\hline
$\Chi^2_{GEV}$ & 11.99 & 23.40 & 28.17 & 14.48 & 21.32 & \textbf{NaN }& 23.26  & 30.90 & \textbf{37.33} & 31.88
\end{tabular}\;.\\[.5em]
\end{center}
Fett hervorgehoben sind die Prüfgrößen, welche größer als die Testgröße sind und die Nullhypothese abgelehnt werden muss.
Für den Landkreis Neuwied liegen lediglich für $1\%$ aller Datensätze (dies entspricht insgesamt $13$ Fällen) das Meldedatum und der Erkrankungsbeginn vor, weshalb keine sinnvolle Auswertung möglich ist. Für den Westerwaldkreis liegt die Prüfgröße nur knapp außerhalb des Konfidenzintervalles.

\subsection{Rangsummentest}

Um zu überprüfen, ob die Mittelwerte der Meldeverzüge der einzelnen Landkreise größer oder kleiner als der Durchschnitt sind, d.h. ob im Schnitt langsamer oder schneller gemeldet wird, würde man eigentlich einen $2$--Stichproben--$t$--Test erwarten. Jedoch gibt es hier das Problem, dass die Voraussetzung derselben Varianz zwischen den Stichproben nicht gegeben ist. Daher müsste auf einen Welch--Test zurückgegriffen werden, welcher jedoch seinerseits eine (annährende) Normalverteilung der Daten voraussetzt. Wenn man davon ausgeht, dass die Meldeverzüge der verschiedenen Landkreise unabhängig voneinander sind, so sollte ein Wilcoxon--Mann--Whitney-Test ($U$--Test, Rangsummentest) durchgeführt werden. Die jeweils einseitige Prüfgröße $U^\pm=\pm 1.65$ für ein Signifikanzniveau von 5\% entspricht den entsprechenden Quantilen der Standardnormalverteilung. Ist die Testgröße $U_{LK}$ der Landkreis--Daten größer als $U^+$, so meldet der entsprechende Landkreis im Schnitt signifikant schneller. Ist hingegen $U_{LK}<U^-$, so meldet der Landkreis im Schnitt signifikant langsamer. Die Testgrößen der einzelnen Landkreise sind:
\begin{center}
\begin{tabular}{c||c|c|c|c|c|c|c|c|c}
Region & KO   & MYK & AW & AK & COC & NR & SIM & EMS & WW \\\hline
$U_{LK}$  & \textbf{1.81} &0.03 & \textbf{-1.93} & \textbf{5.73} & -0.28 & NaN & 0.46 & -1.56 & \textbf{-3.12}
\end{tabular}\;.\\[.5em]
\end{center}
Fett hervorgehoben sind die signifikanten Testergebnisse. Die Stadt Koblenz sowie der Landkreis Altenkirchen meldet also signifikant schneller als der rheinland--pfälzische Durchschnitt, Ahrweiler und der Westerwaldkreis signifikant langsamer.

\begin{figure}[h]
   \hspace{-2.30cm}
    \includegraphics[width=1.241\textwidth]{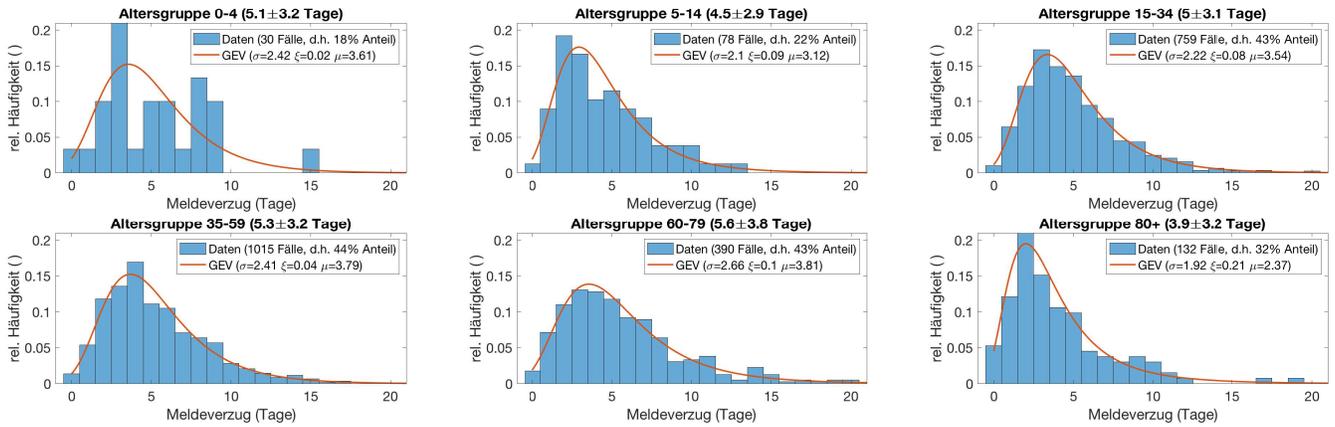} 
    \caption{Meldeverzüge nach Altersklassen. Dargestellt ist jeweils das Histogramm für den Meldeverzug sowie eine angepasste Wahrscheinlichkeitsverteilung (verallgemeinerte Extremwertverteilung / GEV--distribution, siehe \ref{A:Details}). Mittelwerte plus/minus Stichprobenstandardabweichungen der Meldeverzüge sind, ebenso wie der Anteil der Meldeverzüge an der Gesamtzahl der Datensätze, in Klammern angegeben.}
    \label{F:GEV_Age}
\end{figure}

Denselben Test kann man auch auf die mittlere Schnelligkeit in der Meldung nach Altersklassen durchführen. Die einzelnen Testgrößen $U_{A}$ hierfür sind:
\begin{center}
\begin{tabular}{c||c|c|c|c|c|c}
Alter  & 0-4 & 5-14 & 15-34 & 35-59 & 60-79 & 80+  \\\hline
$U_{A}$ & 0.11 & \textbf{1.69} &0.90 & -1.56 & \textbf{-2.05}  & \textbf{ 4.98}
\end{tabular} \; .\\[.5em]  
\end{center}
Hierbei ist festzustellen, dass die Fälle bei 5-14-Jährigen sowie über 80-Jährigen signifikant schneller gemeldet werden als der Durchschnitt, bei 60-80-Jährigen jedoch signifikant langsamer.
Abbildung \ref{F:GEV_Age} liefert eine dazugehörige grafische Repräsentation.

\newpage

\section{Reproduktionszahlen für die einzelnen Landkreise}
\subsection{Koblenz}
\begin{figure}[h]
    \centering
    \includegraphics[width=\textwidth]{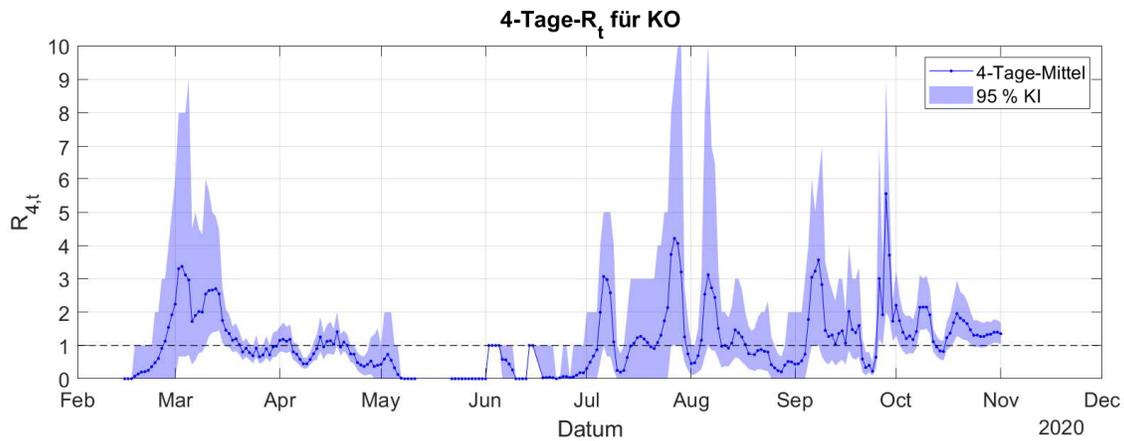}
    \caption{Reproduktionszahl $R_{4,t}$, gemittelt über $4$ Tage, für Koblenz. Dargestellt ist der Mittelwert sowie das $95\%$--Konfidenzintervall basierend auf $1000$ Realisierungen der Schätzung des Meldeverzugs.}
    \label{F:R4-KO}
\end{figure}
\begin{figure}[h]
    \centering
    \includegraphics[width=\textwidth]{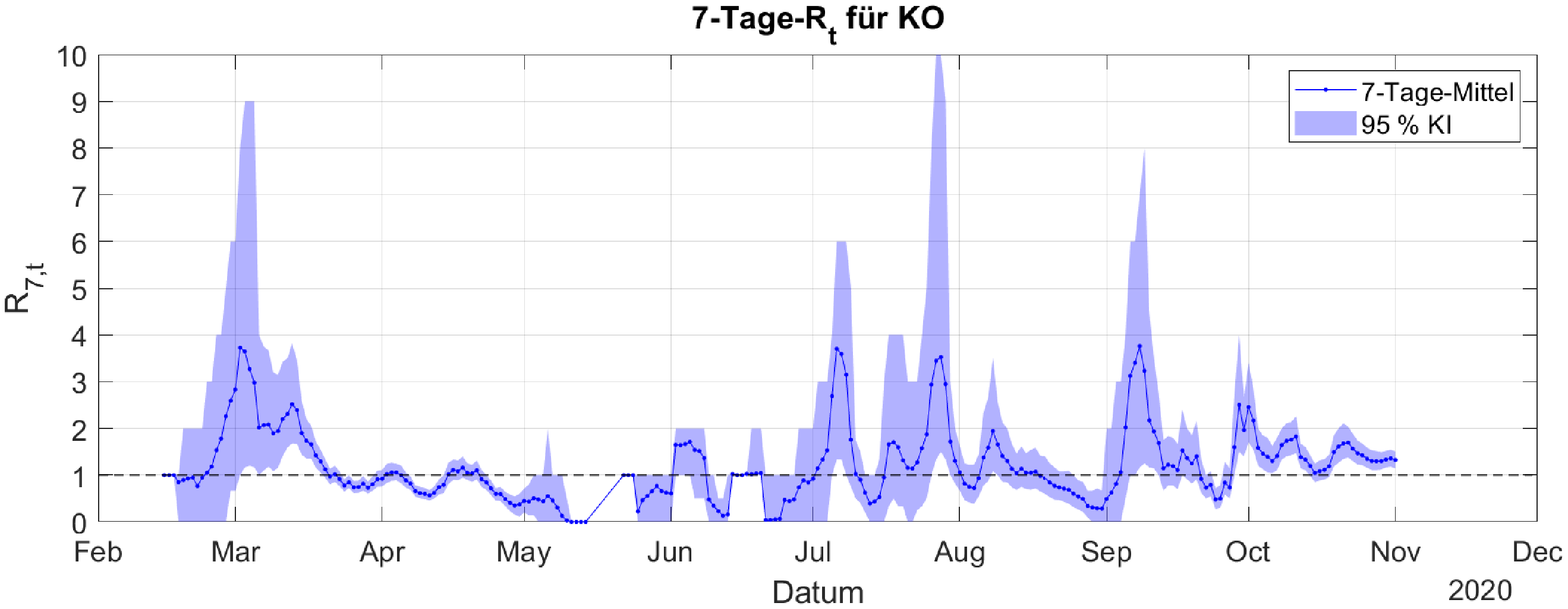}
    \caption{Reproduktionszahl $R_{7,t}$, gemittelt über $7$ Tage, für Koblenz. Dargestellt ist der Mittelwert sowie das $95\%$--Konfidenzintervall basierend auf $1000$ Realisierungen der Schätzung des Meldeverzugs.}
    \label{F:R7-KO}
\end{figure}
\newpage
\subsection{Mayen-Koblenz}
\begin{figure}[h]
    \centering
    \includegraphics[width=\textwidth]{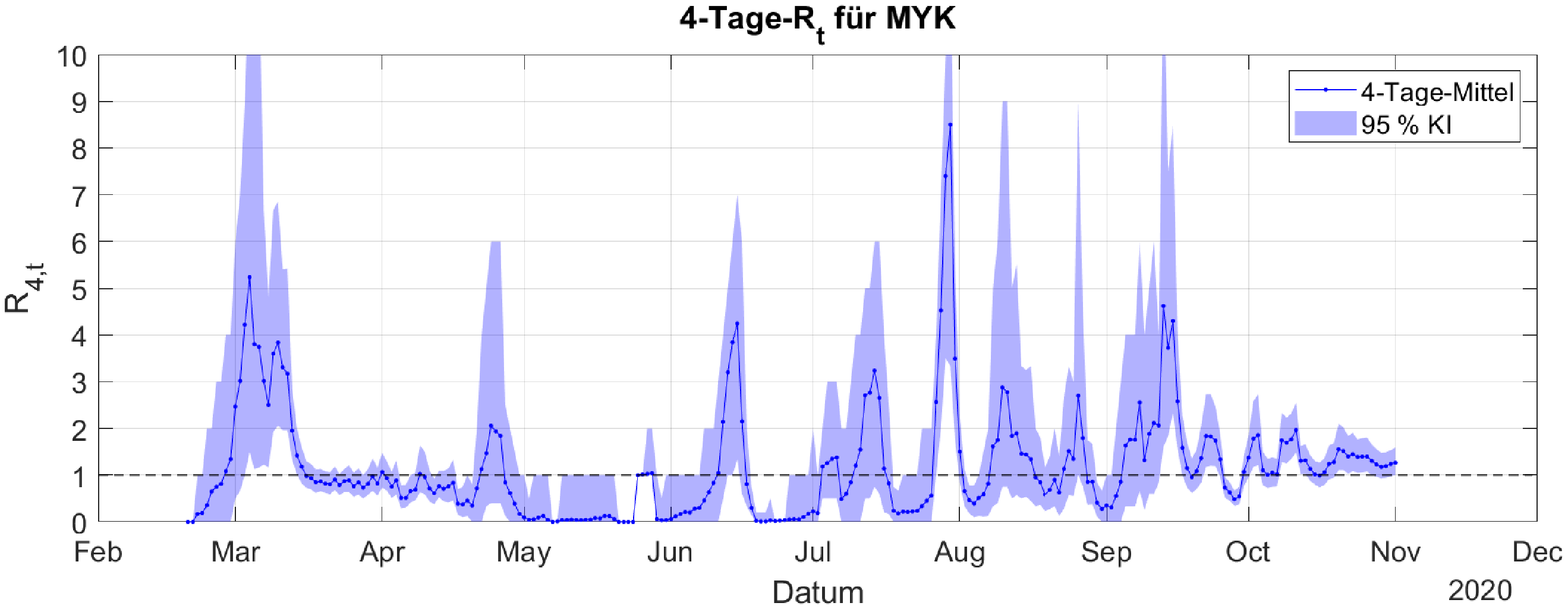}
    \caption{Reproduktionszahl $R_{4,t}$, gemittelt über $4$ Tage, für Mayen--Koblenz. Dargestellt ist der Mittelwert sowie das $95\%$--Konfidenzintervall basierend auf $1000$ Realisierungen der Schätzung des Meldeverzugs.}
    \label{F:R4-MYK}
\end{figure}
\begin{figure}[h]
    \centering
    \includegraphics[width=\textwidth]{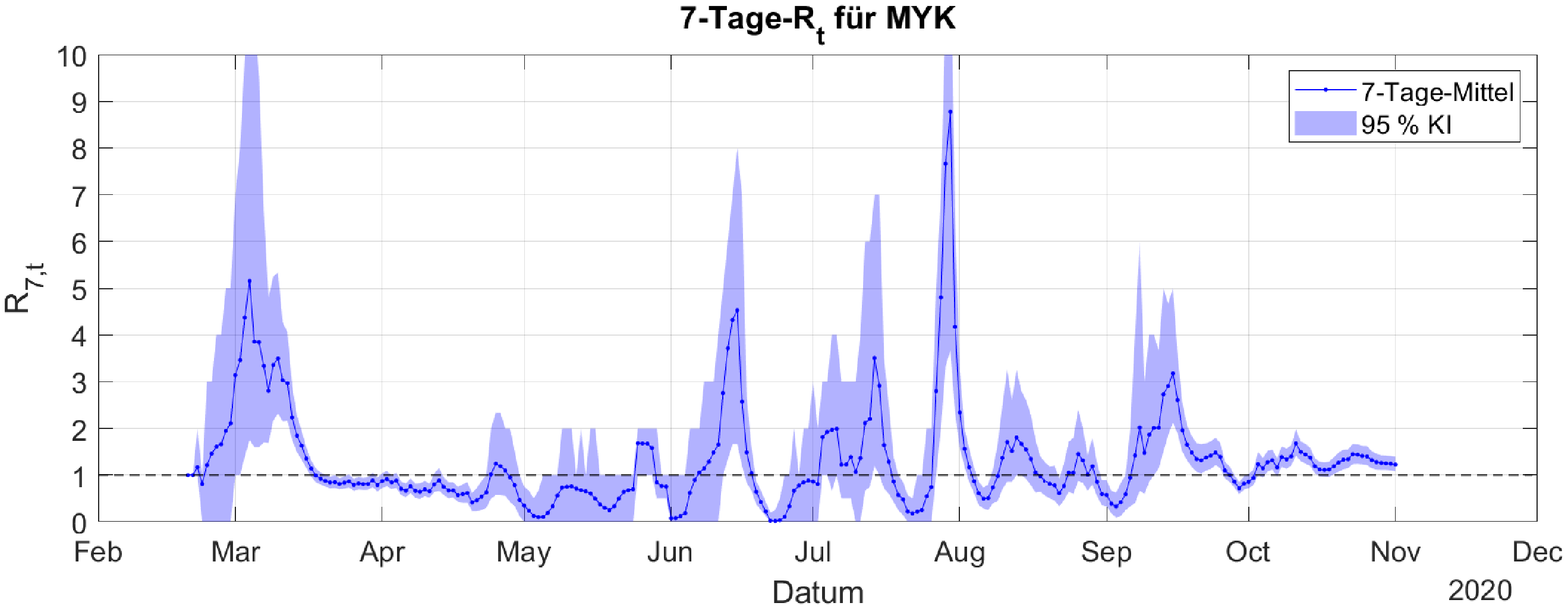}
    \caption{Reproduktionszahl $R_{7,t}$, gemittelt über $7$ Tage, für Mayen--Koblenz. Dargestellt ist der Mittelwert sowie das $95\%$--Konfidenzintervall basierend auf $1000$ Realisierungen der Schätzung des Meldeverzugs.}
    \label{F:R7-MYK}
\end{figure}
\newpage
\subsection{Altenkirchen}
\begin{figure}[h]
    \centering
    \includegraphics[width=\textwidth]{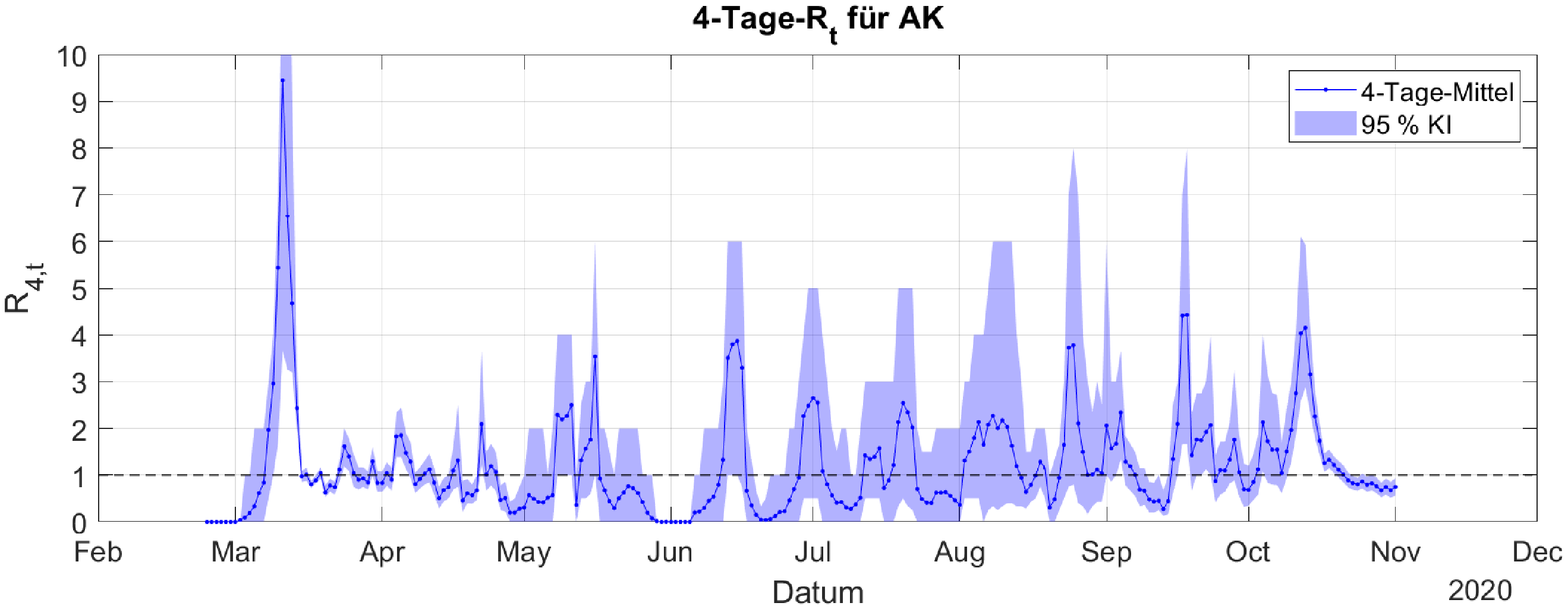}
    \caption{Reproduktionszahl $R_{4,t}$, gemittelt über $4$ Tage, für Altenkirchen. Dargestellt ist der Mittelwert sowie das $95\%$--Konfidenzintervall basierend auf $1000$ Realisierungen der Schätzung des Meldeverzugs.}
    \label{F:R4-AK}
\end{figure}
\begin{figure}[h]
    \centering
    \includegraphics[width=\textwidth]{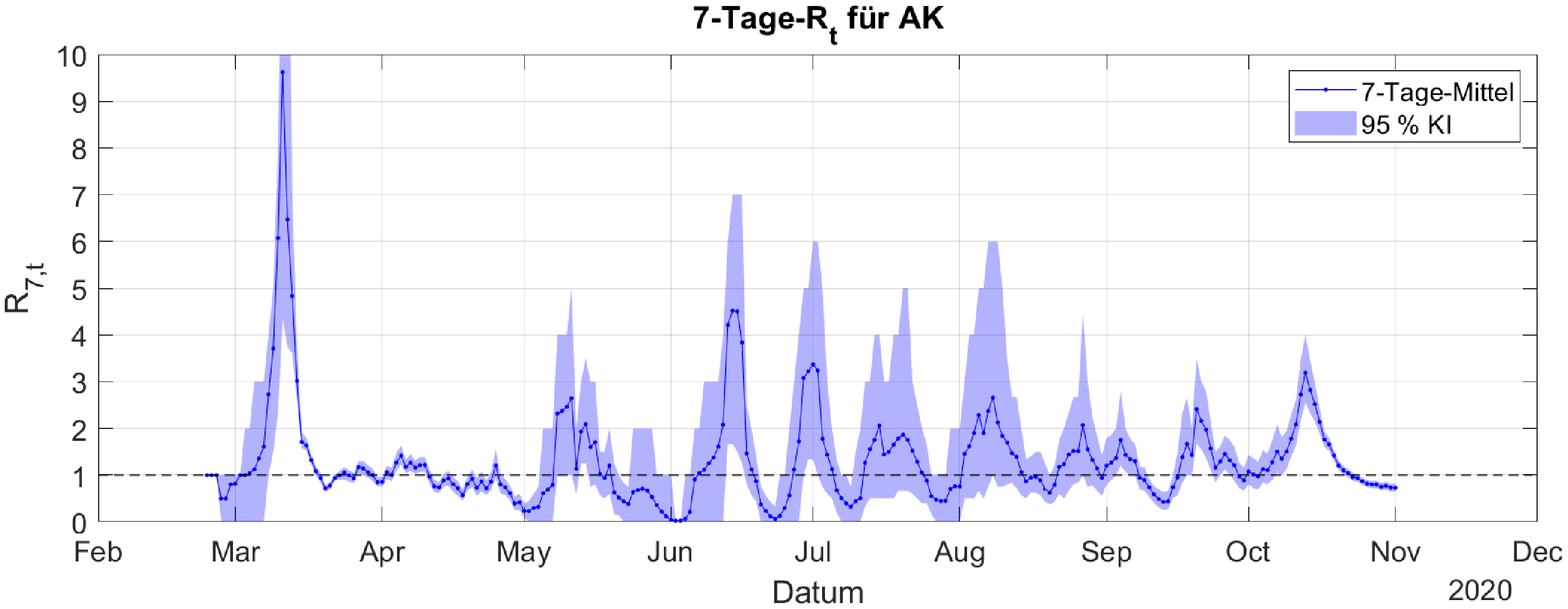}
    \caption{Reproduktionszahl $R_{7,t}$, gemittelt über $7$ Tage, für Altenkirchen. Dargestellt ist der Mittelwert sowie das $95\%$--Konfidenzintervall basierend auf $1000$ Realisierungen der Schätzung des Meldeverzugs.}
    \label{F:R7-AK}
\end{figure}
\newpage
\subsection{Ahrweiler}
\begin{figure}[h]
    \centering
    \includegraphics[width=\textwidth]{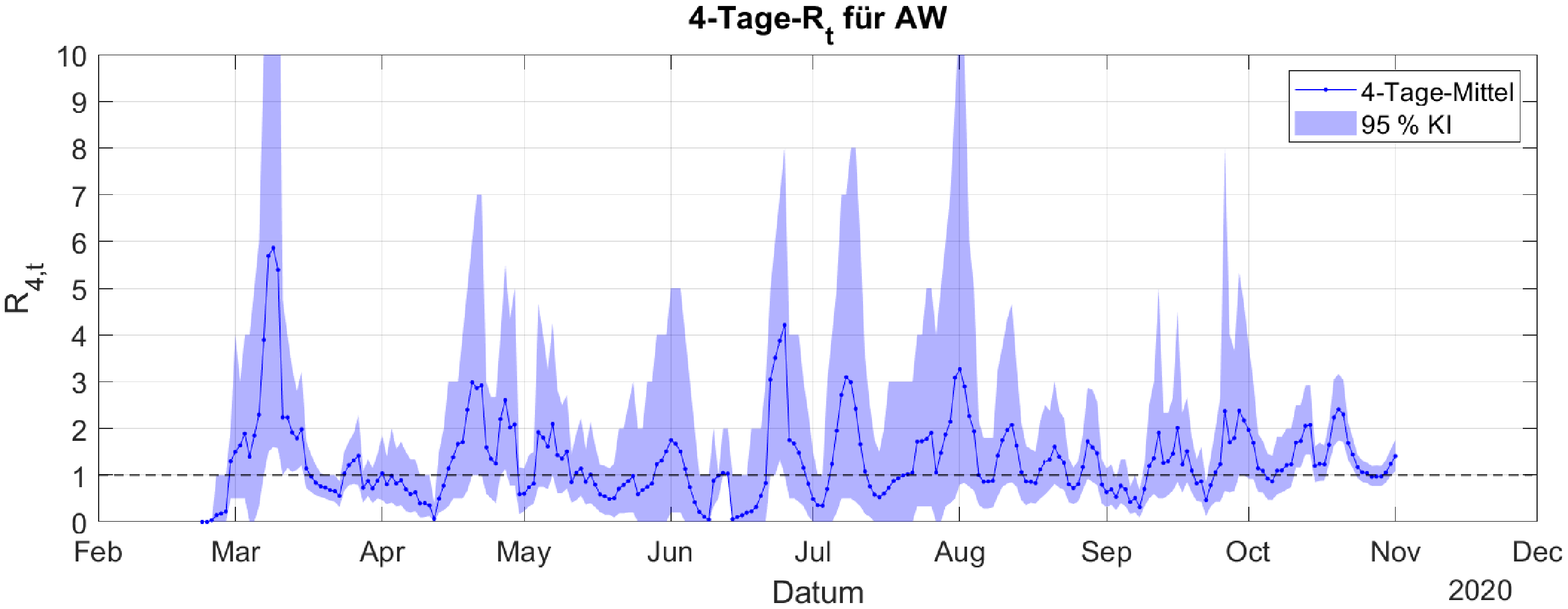}
    \caption{Reproduktionszahl $R_{4,t}$, gemittelt über $4$ Tage, für Ahrweiler. Dargestellt ist der Mittelwert sowie das $95\%$--Konfidenzintervall basierend auf $1000$ Realisierungen der Schätzung des Meldeverzugs.}
    \label{F:R4-AW}
\end{figure}
\begin{figure}[h]
    \centering
    \includegraphics[width=\textwidth]{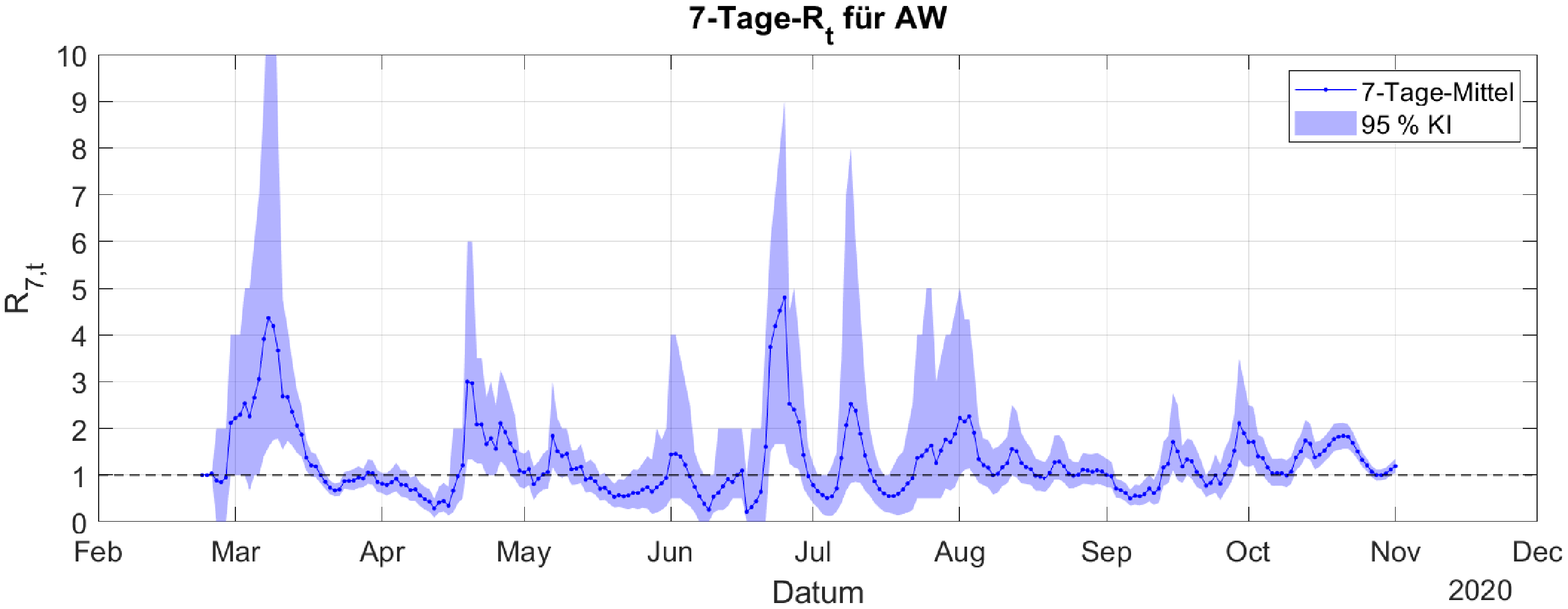}
    \caption{Reproduktionszahl $R_{7,t}$, gemittelt über $7$ Tage, für Ahrweiler. Dargestellt ist der Mittelwert sowie das $95\%$--Konfidenzintervall basierend auf $1000$ Realisierungen der Schätzung des Meldeverzugs.}
    \label{F:R7-AW}
\end{figure}
\newpage
\subsection{Cochem-Zell}
\begin{figure}[h]
    \centering
    \includegraphics[width=\textwidth]{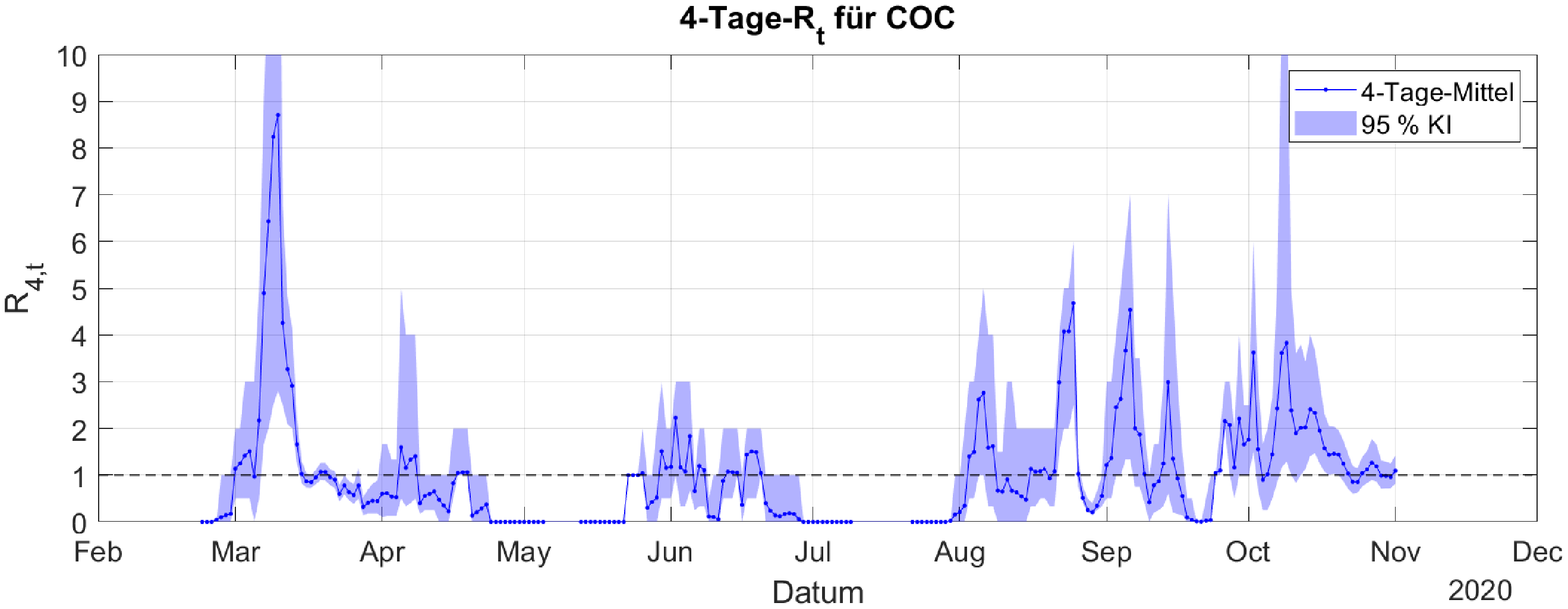}
    \caption{Reproduktionszahl $R_{4,t}$, gemittelt über $4$ Tage, für Cochem--Zell. Dargestellt ist der Mittelwert sowie das $95\%$--Konfidenzintervall basierend auf $1000$ Realisierungen der Schätzung des Meldeverzugs.}
    \label{F:R4-COC}
\end{figure}
\begin{figure}[h]
    \centering
    \includegraphics[width=\textwidth]{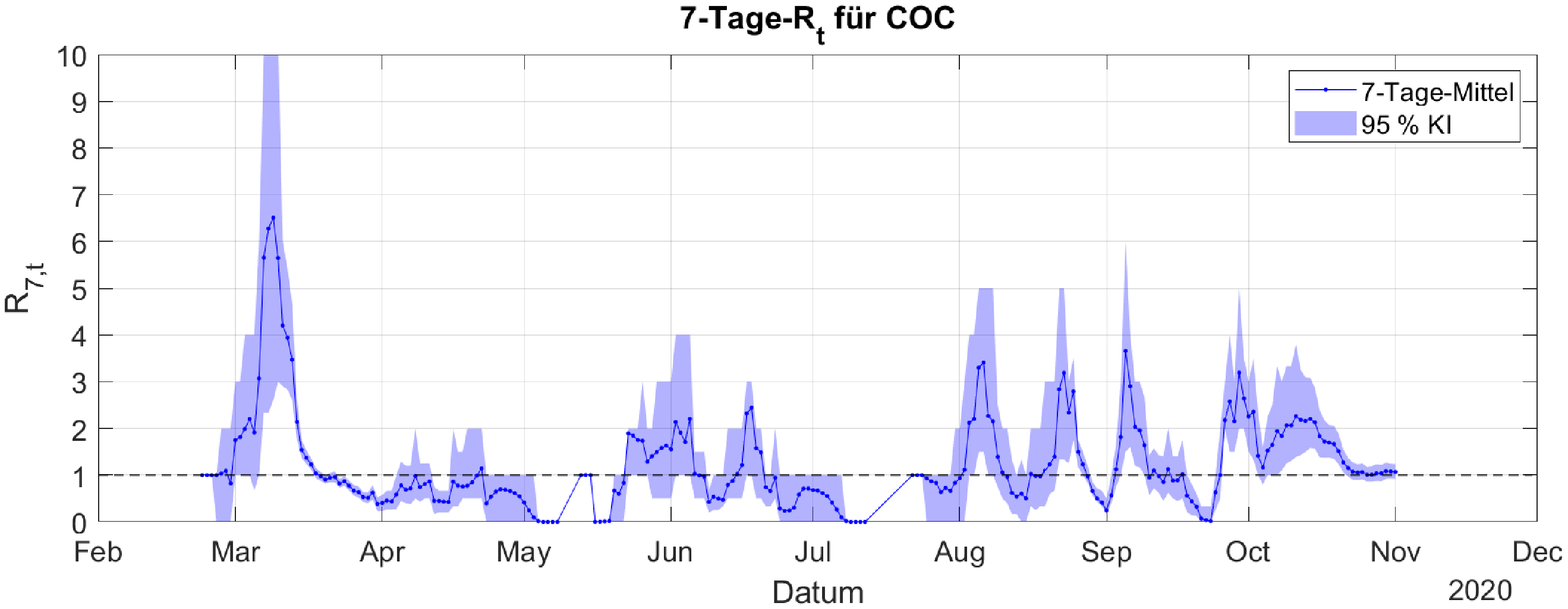}
    \caption{Reproduktionszahl $R_{7,t}$, gemittelt über $7$ Tage, für Cochem--Zell. Dargestellt ist der Mittelwert sowie das $95\%$--Konfidenzintervall basierend auf $1000$ Realisierungen der Schätzung des Meldeverzugs.}
    \label{F:R7-COC}
\end{figure}
\newpage
\subsection{Rhein-Lahn-Kreis}
\begin{figure}[h]
    \centering
    \includegraphics[width=\textwidth]{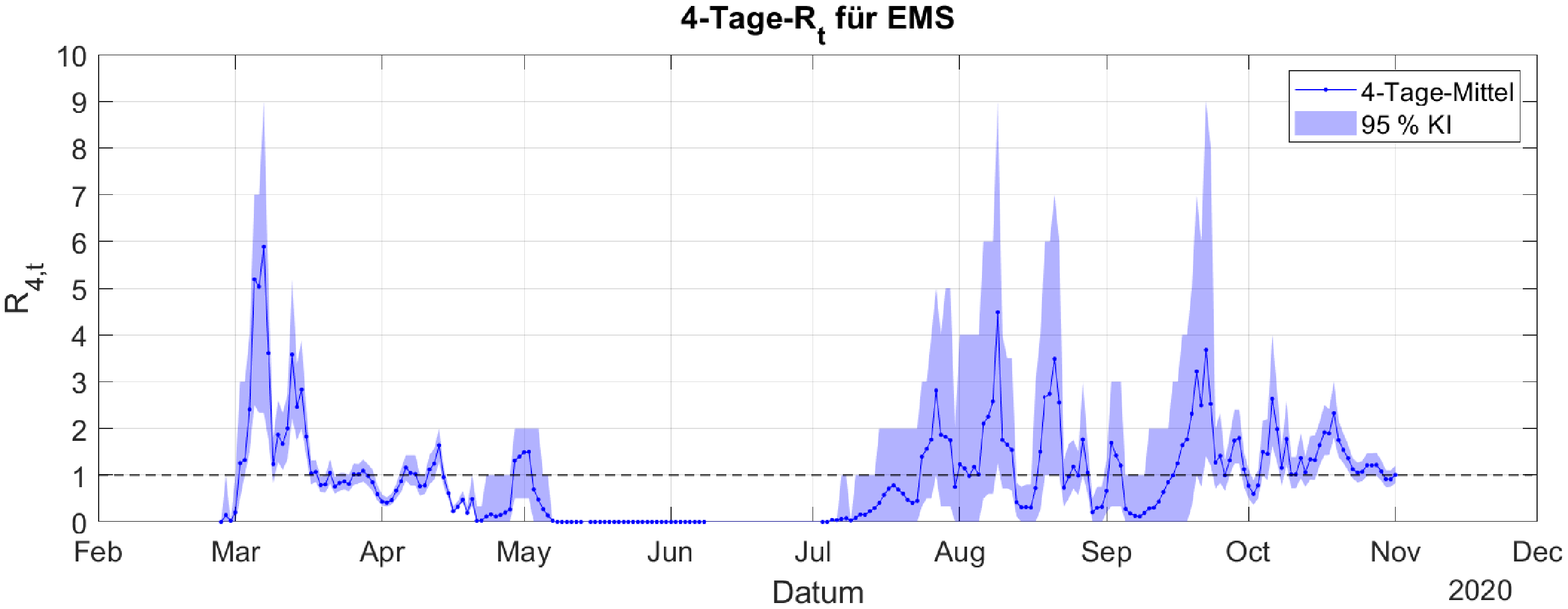}
    \caption{Reproduktionszahl $R_{4,t}$, gemittelt über $4$ Tage, für den Rhein--Lahn--Kreis. Dargestellt ist der Mittelwert sowie das $95\%$--Konfidenzintervall basierend auf $1000$ Realisierungen der Schätzung des Meldeverzugs.}
    \label{F:R4-EMS}
\end{figure}
\begin{figure}[h]
    \centering
    \includegraphics[width=\textwidth]{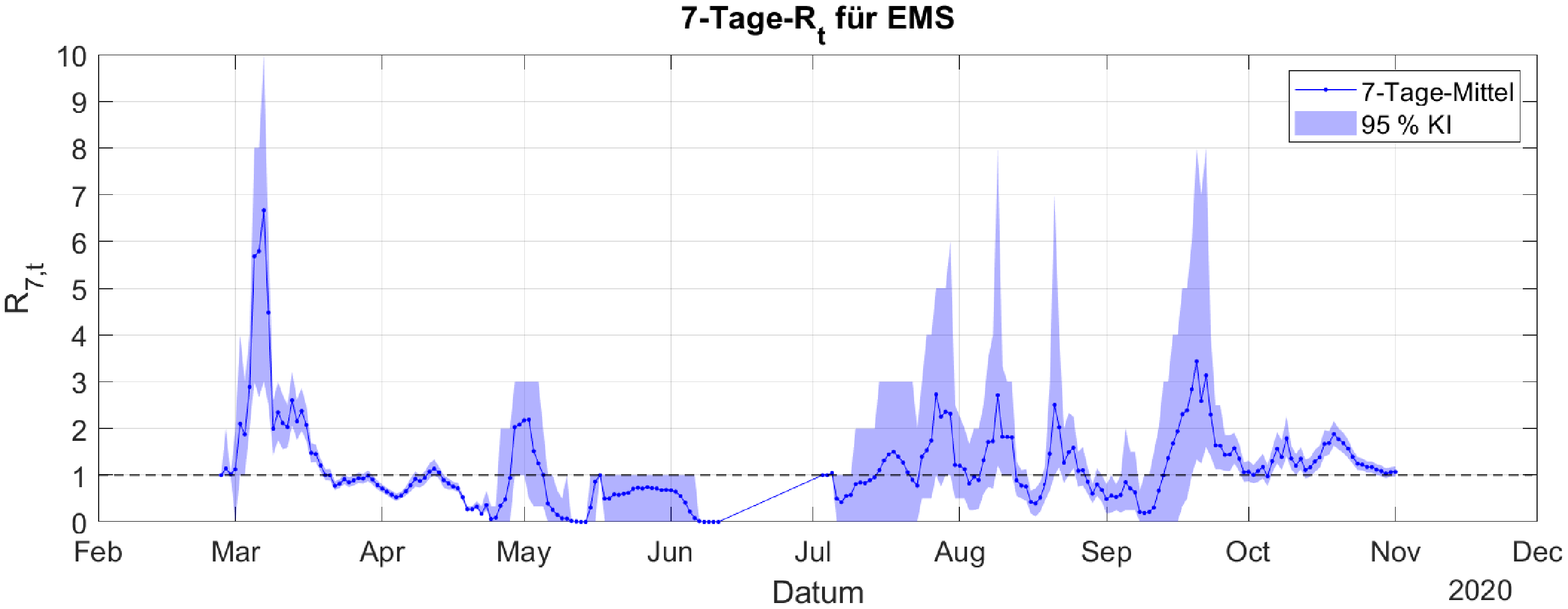}
    \caption{Reproduktionszahl $R_{7,t}$, gemittelt über $7$ Tage, für den Rhein--Lahn--Kreis. Dargestellt ist der Mittelwert sowie das $95\%$--Konfidenzintervall basierend auf $1000$ Realisierungen der Schätzung des Meldeverzugs.}
    \label{F:R7-EMS}
\end{figure}
\newpage
\subsection{Neuwied}
\begin{figure}[h]
    \centering
    \includegraphics[width=\textwidth]{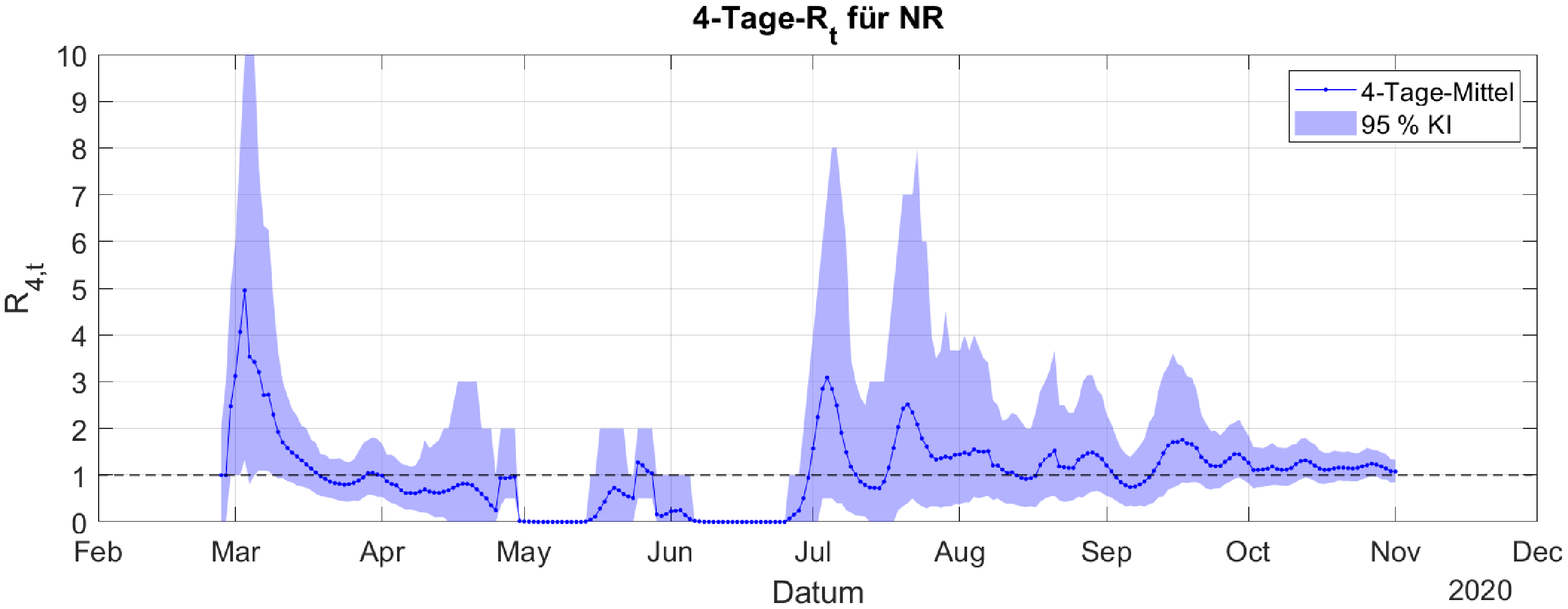}
    \caption{Reproduktionszahl $R_{4,t}$, gemittelt über $4$ Tage, für Neuwied. Dargestellt ist der Mittelwert sowie das $95\%$--Konfidenzintervall basierend auf $1000$ Realisierungen der Schätzung des Meldeverzugs.}
    \label{F:R4-NR}
\end{figure}
\begin{figure}[h]
    \centering
    \includegraphics[width=\textwidth]{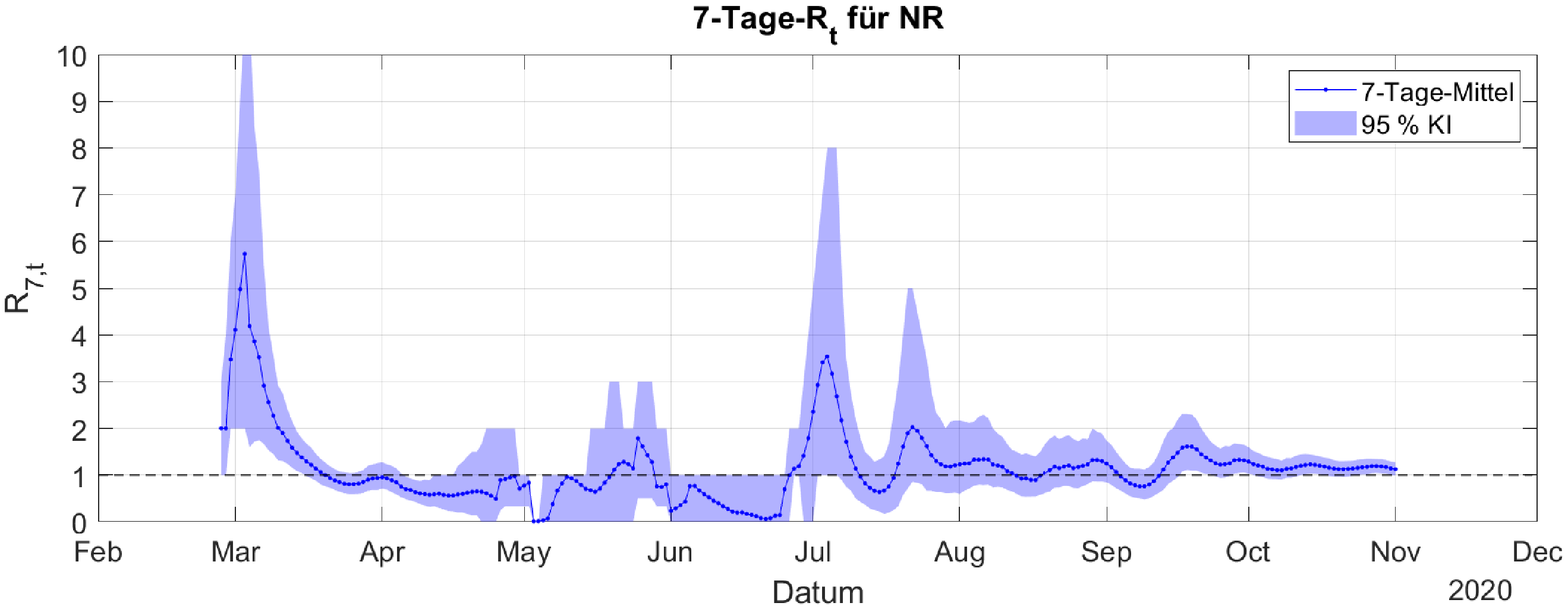}
    \caption{Reproduktionszahl $R_{7,t}$, gemittelt über $7$ Tage, für Neuwied. Dargestellt ist der Mittelwert sowie das $95\%$--Konfidenzintervall basierend auf $1000$ Realisierungen der Schätzung des Meldeverzugs.}
    \label{F:R7-NR}
\end{figure}
\newpage
\subsection{Rhein-Hunsrück-Kreis}
\begin{figure}[h]
    \centering
    \includegraphics[width=\textwidth]{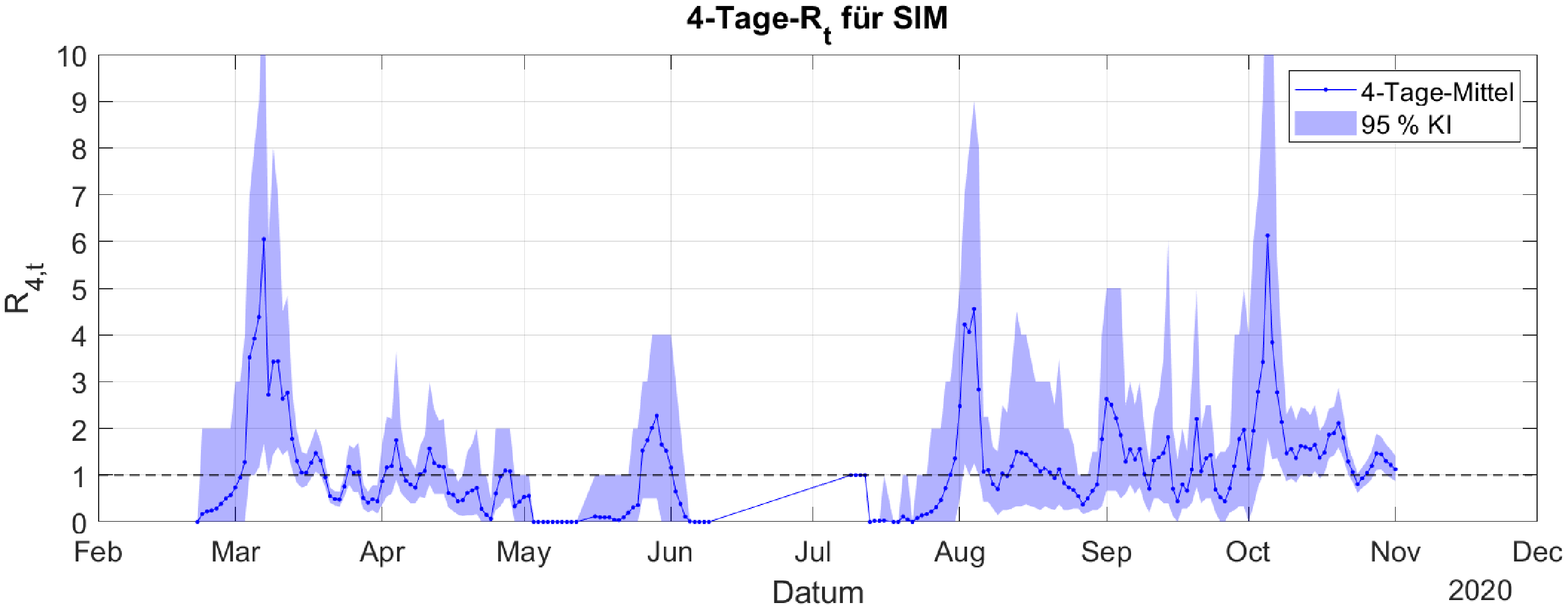}
    \caption{Reproduktionszahl $R_{4,t}$, gemittelt über $4$ Tage, für den Rhein--Hunsrück--Kreis. Dargestellt ist der Mittelwert sowie das $95\%$--Konfidenzintervall basierend auf $1000$ Realisierungen der Schätzung des Meldeverzugs.}
    \label{F:R4-SIM}
\end{figure}
\begin{figure}[h]
    \centering
    \includegraphics[width=\textwidth]{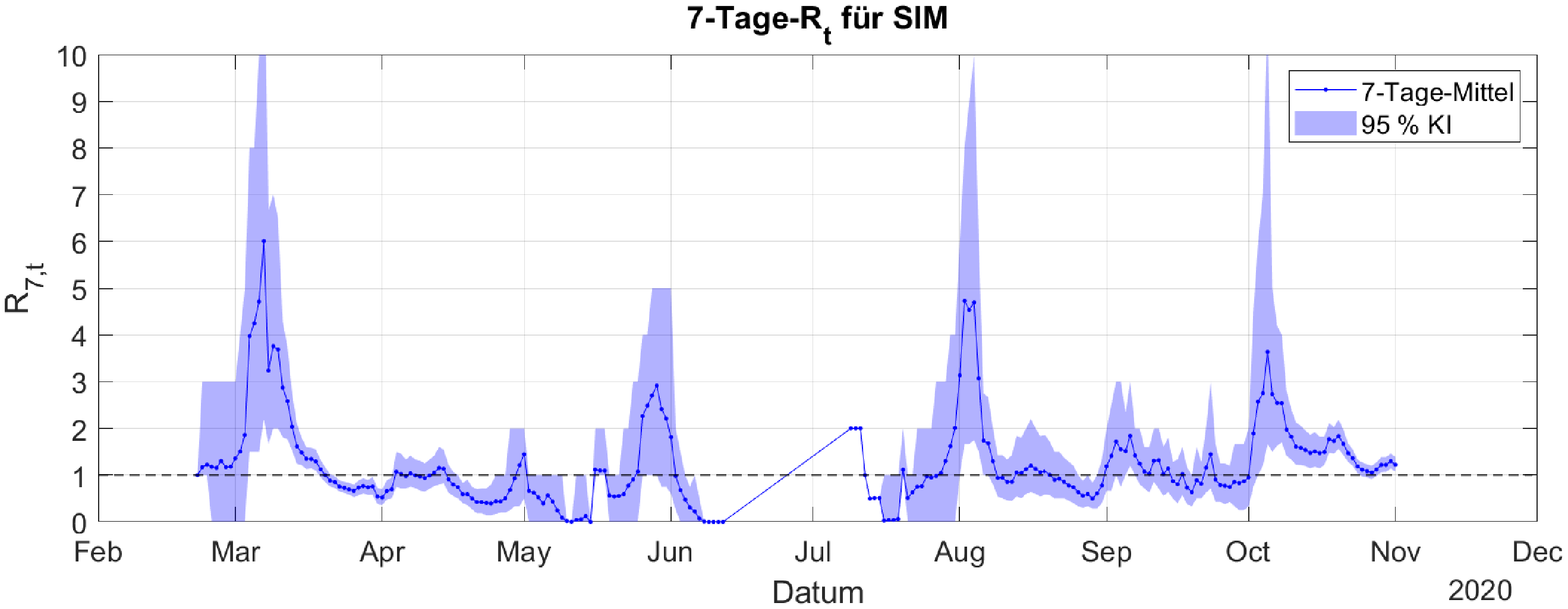}
    \caption{Reproduktionszahl $R_{7,t}$, gemittelt über $7$ Tage, für den Rhein--Hunsrück--Kreis. Dargestellt ist der Mittelwert sowie das $95\%$--Konfidenzintervall basierend auf $1000$ Realisierungen der Schätzung des Meldeverzugs.}
    \label{F:R7-SIM}
\end{figure}
\newpage
\subsection{Westerwaldkreis}
\begin{figure}[h]
    \centering
    \includegraphics[width=\textwidth]{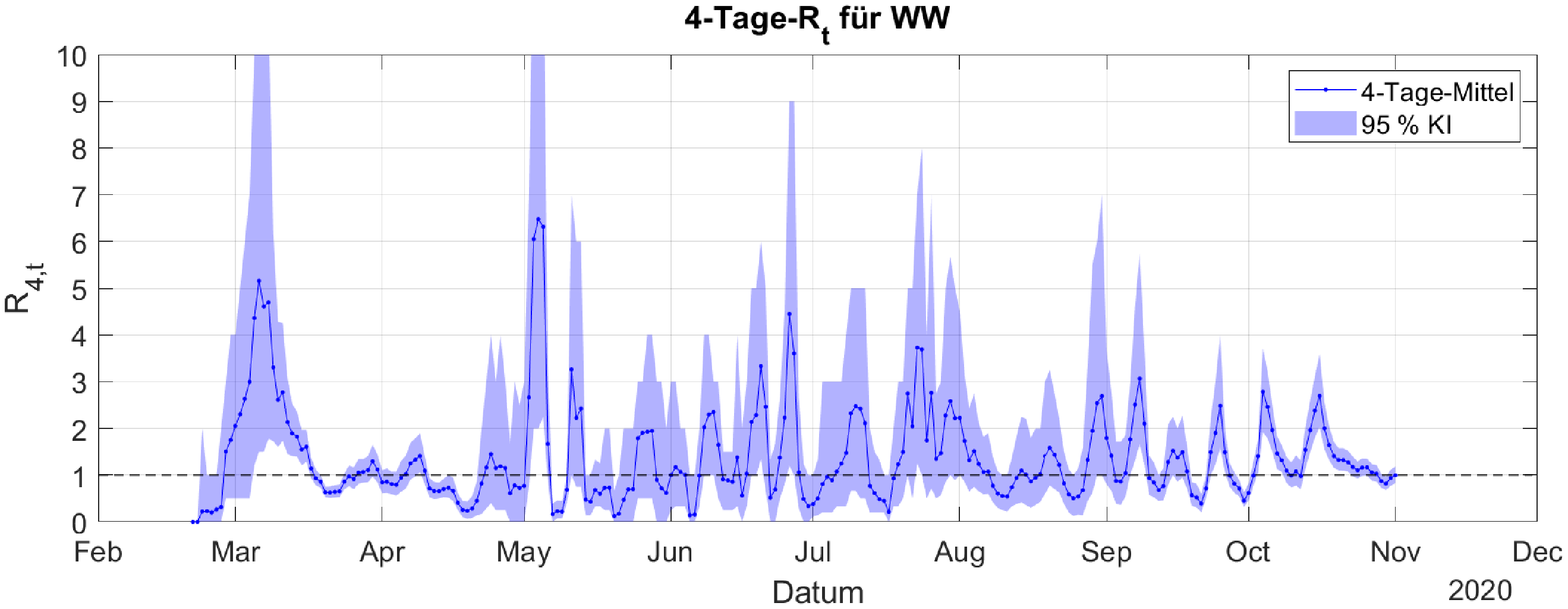}
    \caption{Reproduktionszahl $R_{4,t}$, gemittelt über $4$ Tage, für den Westerwaldkreis. Dargestellt ist der Mittelwert sowie das $95\%$--Konfidenzintervall basierend auf $1000$ Realisierungen der Schätzung des Meldeverzugs.}
    \label{F:R4-WW}
\end{figure}
\begin{figure}[h]
    \centering
    \includegraphics[width=\textwidth]{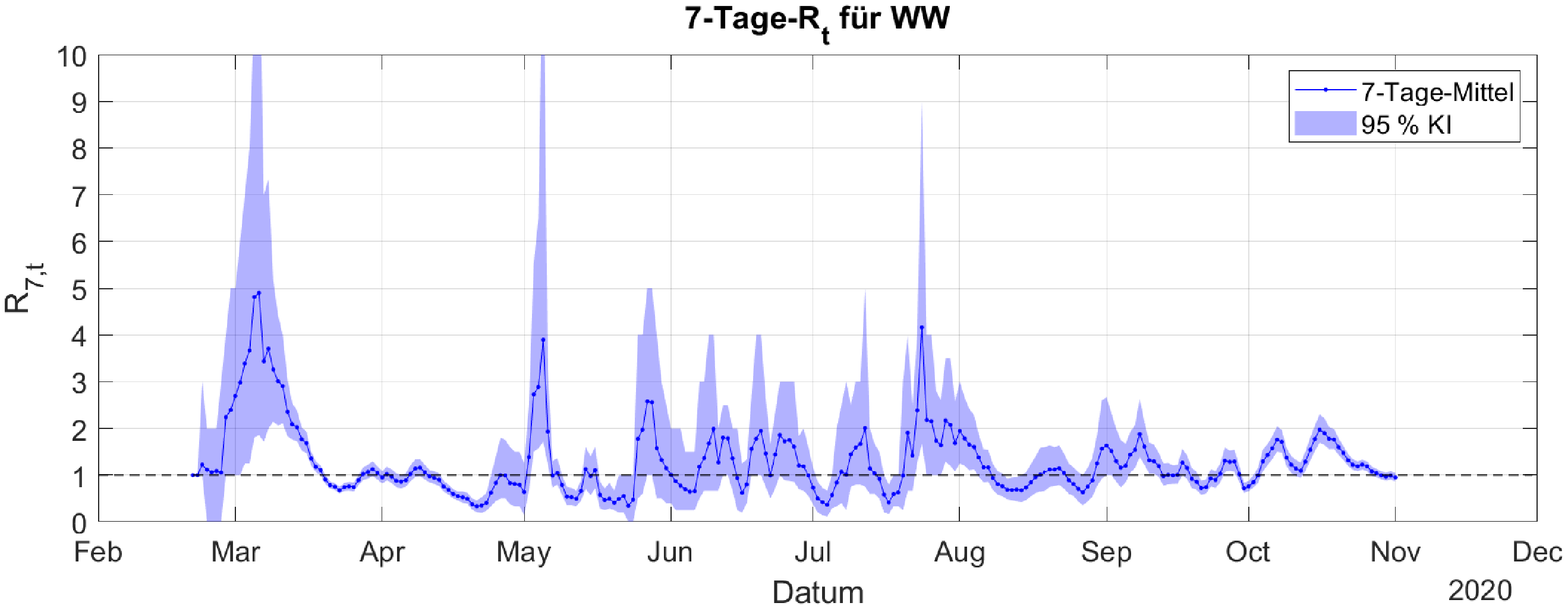}
    \caption{Reproduktionszahl $R_{7,t}$, gemittelt über $7$ Tage, für den Westerwaldkreis. Dargestellt ist der Mittelwert sowie das $95\%$--Konfidenzintervall basierend auf $1000$ Realisierungen der Schätzung des Meldeverzugs.}
    \label{F:R7-WW}
\end{figure}


\newpage
\begin{thebibliography}{99}

\bibitem{COV:RKI} Robert--Koch-Institut, \emph{Fallzahlen in Deutschland}, \url{https://www.arcgis.com/home/item.html?id=f10774f1c63e40168479a1feb6c7ca74}
\bibitem{COV_readme:RKI} Robert--Koch-Institut, \emph{Beschreibung der Daten des RKI Covid-19-Dashboards}, 
\url{https://npgeo-corona-npgeo-de.hub.arcgis.com/datasets/dd4580c810204019a7b8eb3e0b329dd6_0}
\bibitem{EstR:RKI} Robert--Koch--Institut, \emph{Erläuterung der Schätzung der zeitlich variierenden Reproduktionszahl $R$}, Mai 2020, \url{https://www.rki.de/DE/Content/InfAZ/N/Neuartiges_Coronavirus/Projekte_RKI/R-Wert-Erlaeuterung.pdf?__blob=publicationFile}.
\bibitem{GH20} Götz, T., Heidrich, P., \emph{Early stage COVID-19 disease dynamics in Germany: models and parameter identification}. J.Math.Industry 10, 20 (2020). doi: \url{10.1186/s13362-020-00088-y}
\bibitem{CDea20} Contreras, S., Dehning, J., Loidolt, M., Spitzner, F.P., Urrea-Quintero, J.H., Mohr, S.B., Wilczek, M., Zierenberg, J., Wibral, M., Priesemann, V., \emph{The challenges of containning SARS-CoV-2 via test-trace-and-isolate}, September 2020, \url{https://arxiv.org/abs/2009.05732}
\bibitem{Jen55} Jenkinson, A.F., \emph{The frequency distribution of the annual maximum (or minimum) values of meteorological elements}. Q.J.R.~Meteorol.~Soc. 81 (348): 158–171 (1955). doi: \url{10.1002/qj.49708134804} 
\end{thebibliography}
\end{document}